\documentclass[11pt, 3p]{elsarticle}

\usepackage{mathpazo}
\usepackage[T1]{fontenc}
\usepackage[utf8]{inputenc}
\usepackage{multirow}
\usepackage[colorlinks=true]{hyperref}
\usepackage[english]{babel}
\usepackage{adjustbox}
\usepackage{xcolor}
\usepackage{booktabs}
\usepackage{amsmath}
\usepackage{textcomp}
\usepackage{lineno}
\usepackage{graphicx}
\usepackage{tikz}
\usepackage{siunitx}
\usepackage{subcaption}
\usepackage{setspace}
\usepackage{natbib}

\usepackage{caption}
\usepackage[font=small,labelfont=bf, skip=0pt]{caption}
\biboptions{sort&compress}

\usepackage{soulutf8}
\newcommand{\edited}[1]{#1}
\newcommand{\reedited}[1]{#1}

\def\Fig{Fig.\,}
\def\Figs{Figs.\,}
\def\Tab{Table\,}

\def\Ref{Ref.\,}

\biboptions{sort&compress}

\graphicspath{{./}} 

\begin{document}
	
\begin{frontmatter}

\journal{Scripta Materialia}

\title{T2 phase site occupancies in the Cr--Si--B system: a combined synchroton-XRD/first-principles study}
 
\author[1]{Thiago T. Dorini}
\author[2]{Bruno X. de Freitas}
\author[2]{Pedro P. Ferreira}
\author[2]{Nabil Chaia}
\author[2]{Paulo A. Suzuki}
\author[3]{Jean-Marc Joubert}
\author[2]{Carlos A. Nunes}
\author[2]{Gilberto C. Coelho}
\author[2]{Luiz T. F. Eleno\corref{cor}}
\cortext[cor]{Corresponding author: +55 12 3159 9810, website: \url{https://computeel.org}}
\ead{luizeleno@usp.br}

\address[1]{Universit{\'e} de Lorraine, CNRS, IJL, Nancy, France}
\address[2]{Computational Materials Science Group (ComputEEL/MatSci), Escola de Engenharia de Lorena da Universidade de S{\~a}o Paulo (EEL--USP), Materials Engineering Department (Demar). Lorena--SP, Brazil}
\address[3]{Univ. Paris Est Creteil, CNRS, ICMPE, UMR 7182, 2 rue Henri Dunant, 94320 Thiais, France}

\begin{abstract}


	Boron and Silicon site occupancies of the T2 phase in the Cr--Si--B system were investigated experimentally and by first-principles electronic-structure calculations. A sample with nominal composition Cr$_{0.625}$B$_{0.175}$Si$_{0.2}$ was arc-melted under argon, encapsulated in a quartz-tube and heat-treated at 1400°C for 96 hours. It was then analyzed using Scanning Electron Microscopy (SEM) and X-Ray Diffractometry (XRD) with synchrotron radiation. An excellent agreement was obtained between experiments and theoretical calculations, revealing that Si occupies preferably the $4a$ sublattice of the structure, \edited{as boron atoms in $8h$ nearest-neighbors form very stable \emph{dumbbell} bonds. Site preferences are thus a key factor for the stabilization of the structure.} The results of this work provide important information to support a better description of this phase in alloys with Si and B, since T2 phases are known to occur in many important Transition Metal--Si--B ternary systems, such as Nb/Mo/W/Ta/V--Si--B.
	
\end{abstract}

\begin{keyword}
X-ray diffraction (XRD) \sep Density Functional Theory (DFT) \sep Refractory metals \sep Cr-Si-B \sep Intermetallics
\end{keyword}

\end{frontmatter}



\edited{Recent demands for energy generation and consumption require the search for new materials and processing routes capable of stabilizing mechanical and thermal properties during long times under extreme conditions. For example, among the key materials for high temperature applications are multicomponent systems containing refractory metals (RM) such as Nb and Mo, often modified with Si and B \mbox{\cite{Bewlay2003, Santos2019,Ito2001,Ito2003, Schneibel2002, Park2002, Sakidja2005}}, originating the so-called RM-silicides.}
However, these alloys are known to exhibit \edited{low to} moderate resistance to oxidation/corrosion at elevated temperatures. If the modification of the alloy \edited{composition is not effective in enhancing} the oxidation behavior at high temperatures, the deposition of protective coatings is required to ensure the integrity of parts during service \cite{Knittel2013, Portebois2013, Perepezko2010}. Coatings developed in the past for this class of materials are based on silicide compounds that are able to form a silica protective layer during exposure \cite{Knittel2013, Dimiduk2003}. Additions of boron to silicides is of particular interest, leading to the modification of the silica layer by lowering its viscosity, which can help to heal \edited{fatigue cracks induced by thermal cycling}. \edited{Therefore}, knowledge of phase diagrams and the crystal structures associated with intermetallic phases present in RM--Si--B systems is very important to develop new compositions for alloys and coatings.

Among the key systems, Cr--Si--B is a promising candidate, especially alloys in the Cr-rich region \cite{Nowotny1958, Villela2011, Chad2008}. Despite its potential in a wide range of applications, there is a lack of experimental and theoretical data in the literature. \edited{In particular, the Cr--Si--B system} contains a phase, usually denoted by T2 (which is a ternary extension of Cr$_5$B$_3$ with silicon substitution for boron), with very scarce information regarding its site occupancies and enthalpy of formation. 
 
Therefore, the aim of the present work is to determine the sublattice occupancies of the T2 compound using synchrotron X-ray diffraction measurements and Density Functional Theory (DFT) calculations in order to help to improve  the reliability of Cr--Si--B thermodynamic descriptions. In the following discussion, we demonstrate that Si atoms, in fact, have a \edited{preference} for one of the boron sublattices of the T2 phase, in contrast to what was proposed in the optimization performed by \citet{Villela2011}. That information calls for a new, appropriate thermodynamic reassessment of the Cr--Si--B system. \edited{In addition, besides providing invaluable information on the poorly-studied Cr--Si--B system, our work also sheds light on the need for better descriptions of related systems}, as T2 phases are known to occur in many important ternaries, such as Nb/Mo/W/Ta/V--Si--B \cite{Sun2011, Sakidja2008, Colinet2014, Rodrigues2007, Candioto2001, Mendiratta2002, Parthasarathy2002, Zhang2013, Nabil2017, rodrigues2004lattice, Khan2012, Nunes2009, Nunes2012}.

\begin{figure*}
	\centering
	\begin{minipage}{.49\textwidth}
		\centering
      \includegraphics[width=.8\textwidth]{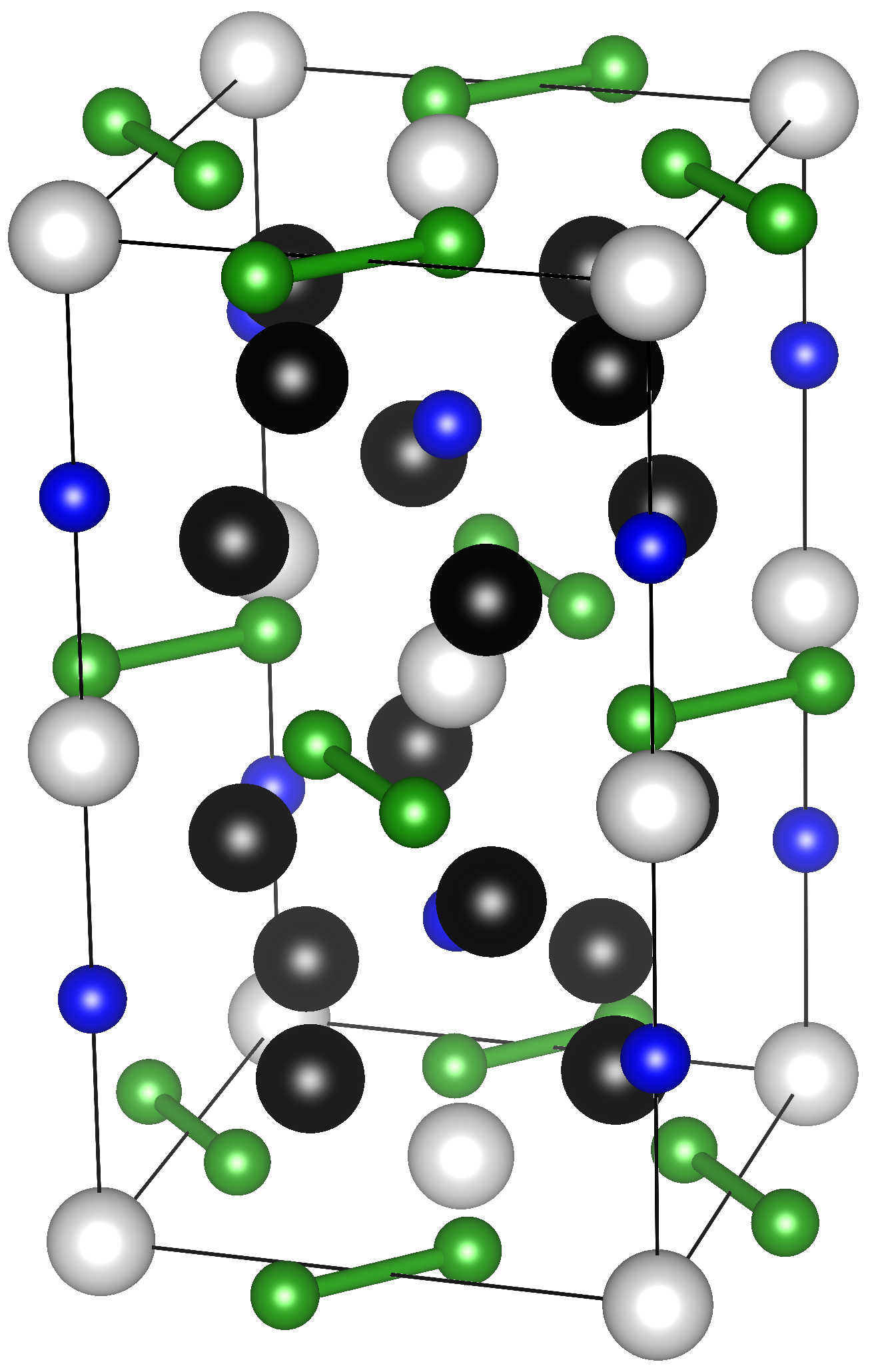}
		\end{minipage}
		\begin{minipage}{.49\textwidth}
			\centering
			\includegraphics[width=.8\textwidth]{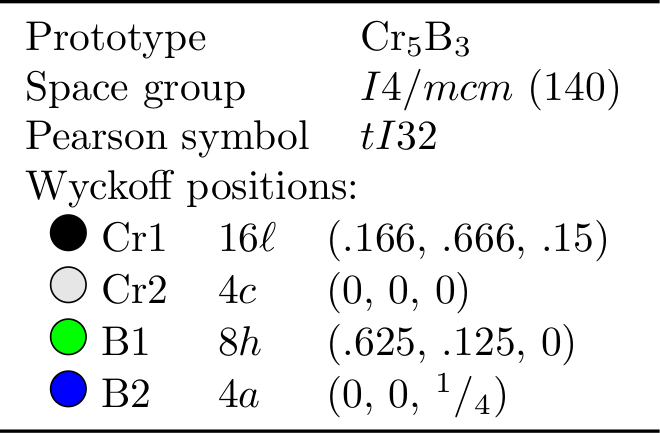}
		\end{minipage}
    \caption{(a) Crystallographic information on the T2--Cr$_5$B$_3$ compound \cite{Gianoglio1983, Villars1995}. \edited{The only bonds shown are the \emph{dumbbells} between two nearest-neighbor boron ($8h$) atoms.}}
\label{fig:tab:structure-t2}
\end{figure*}


The tetragonal unit cell and crystallographic information on the T2--Cr$_5$B$_3$ compound  \cite{Gianoglio1983, Villars1995} are presented in \Fig \ref{fig:tab:structure-t2}. Cr atoms ocuppy the $16\ell$ and $4c$ sites, while B occupies the $8h$ and $4a$ Wyckoff positions. From previous microanalysis experiments \cite{Villela2011}, Si dissolves in this boride by replacing B atoms, keeping the chromium content constant \edited{at a molar fraction of} $x_\text{Cr}=0.625$. A sample was then prepared with nominal composition Cr$_{0.625}$B$_{0.175}$Si$_{0.2}$. It was obtained by arc melting of high-purity Cr (min. 99.9 wt\%), Si (min. 99.99 wt\%) and B (min 99.5 wt\%) in a water-cooled copper crucible under analytical argon (min 99.995\%) atmosphere. After the melting process, \edited{weight} losses were lower than 1.3\%. The sample was then encapsulated in a quartz tube under vacuum and heat-treated at 1400°C for 96h in a furnace with constant flow of analytical argon. 
The sample was characterized at room temperature through Scanning Electron Microscopy/Back-Scattered Electron Images (SEM/BSE) \edited{in the as-cast and heat-treated  states}, the latter being also submitted to XRD from synchrotron radiation. SEM/BSE micrographs were obtained from flat and polished specimens and XRD experiments were carried out in high resolution mode using a multiple axes Huber diffractometer located at the Brazilian Synchroton Light Laboratotry (LNLS) in Campinas (SP), Brazil. To improve randomness of orientation, the powder sample, sieved to below $53\,\si{\micro m}$, was placed in a rotating 5\,mm diameter--1\,mm depth cylindrical support. The measurement was performed with a monochromatic X-ray beam ($\lambda$ = 1.237\,\AA). 
\reedited{The combination of flat plate geometry and parallel beam gives results with a high peak/noise ratio, accurate enough to provide site occupancies with small standard deviations.}
Phase identification and refinement were performed using the Rietveld method \cite{Rietveld1968, Rodriguez1990} with input crystallographic data from the literature \cite{Gianoglio1983, Kuzma1982, Okada1987}.

\begin{figure*}
	\centering
	\begin{subfigure}{0.5\textwidth}
		\centering
        \includegraphics[width=.99\textwidth]{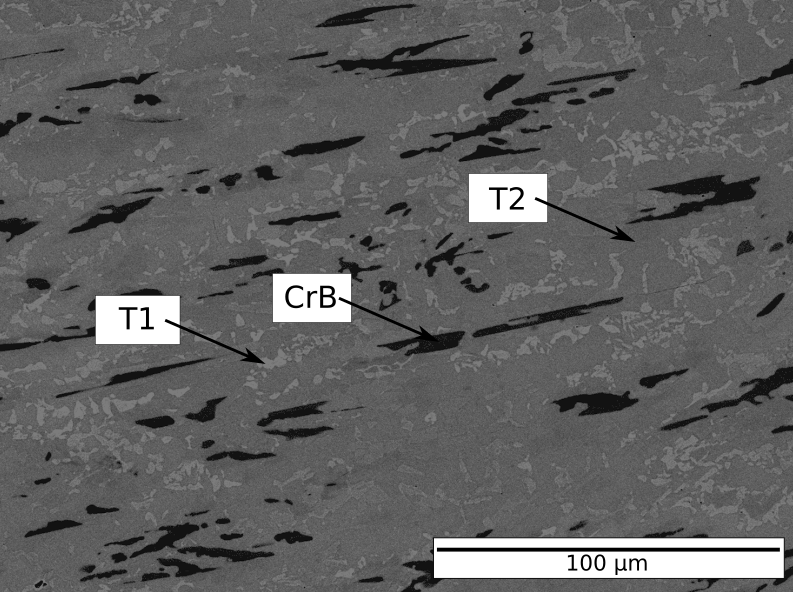}
		\caption{As cast.}
	\end{subfigure}%
	\begin{subfigure}{.5\textwidth}
		\centering
        \includegraphics[width=.99\textwidth]{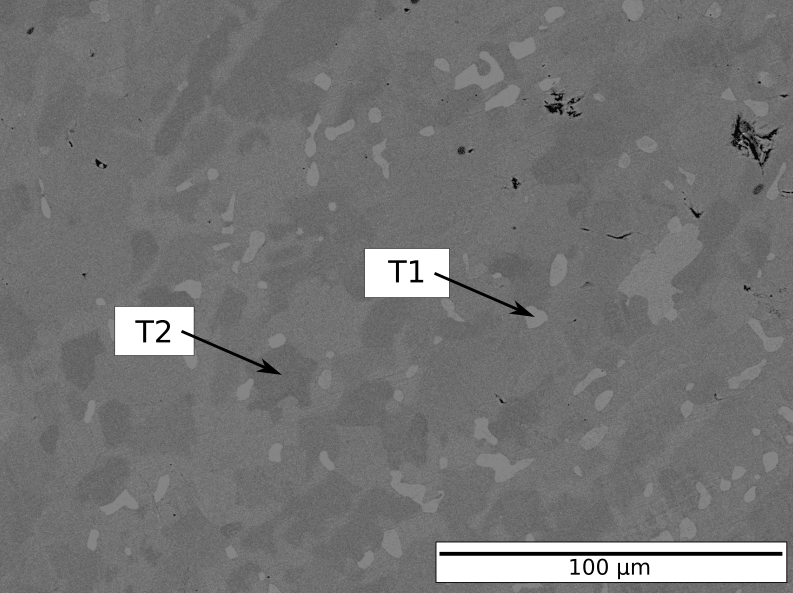}
		\caption{Heat treated at 1400\textdegree{}C for 96h.}
	\end{subfigure}
    \caption{\edited{SEM images in BSE contrast} of the Cr$_{0.625}$B$_{0.175}$Si$_{0.2}$ sample, showing (a) the presence of CrB and T1--Cr$_5$Si$_3$ precipitates in a T2--Cr$_5$(Si,B)$_3$ matrix in the as-cast state, and (b) the microstructure after heat treatment at 1400°C for 96\,h, with a two-phase microstructure constituted by T1 and T2.}
	\label{fig:mev-4} 
\end{figure*}

\begin{figure}[t]
	\centering
	\includegraphics[width=.9\columnwidth]{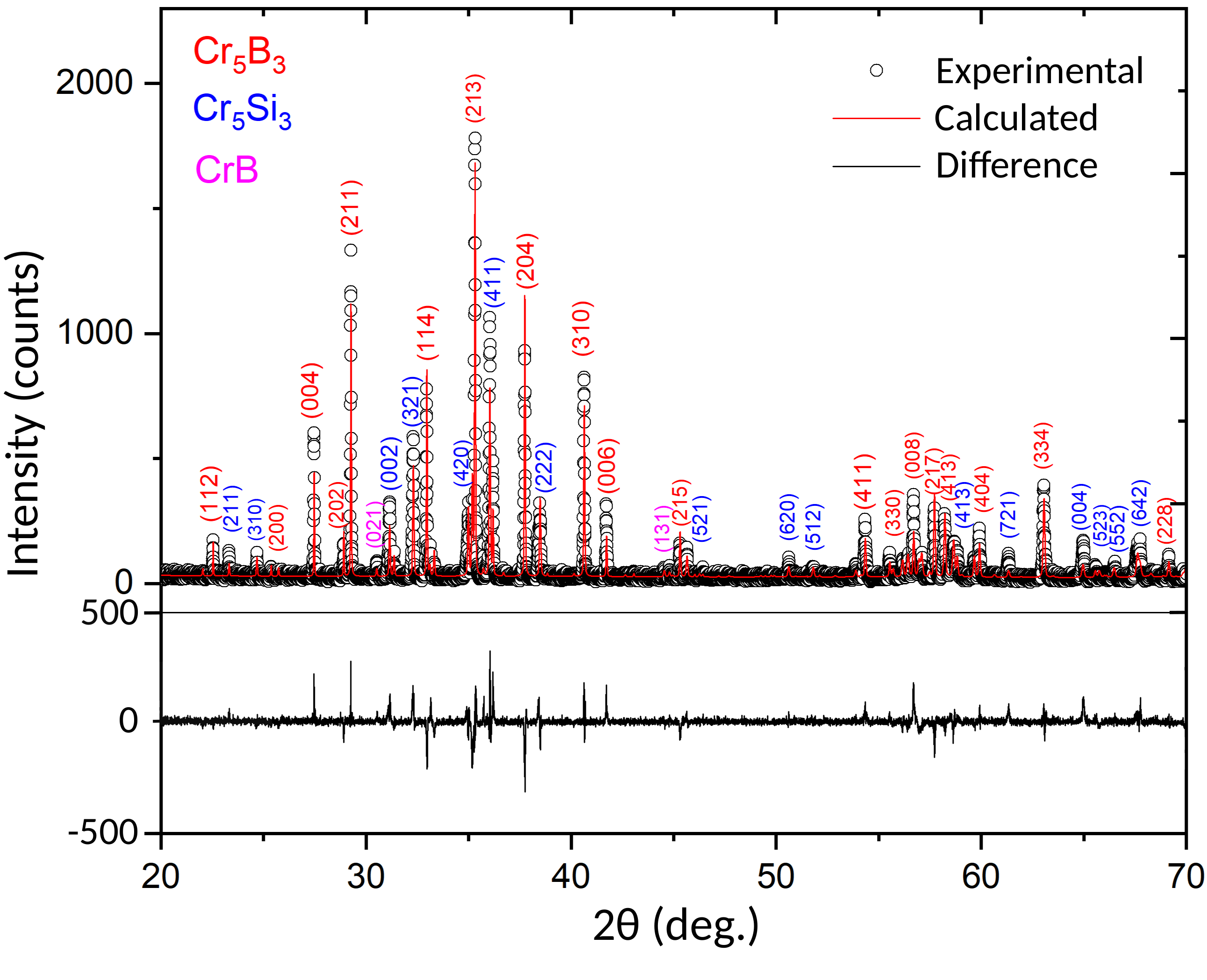}
    \caption{Synchrotron x-ray diffraction pattern of the Cr$_{0.625}$B$_{0.175}$Si$_{0.2}$ sample heat-treated at 1400°C for 96 hours, showing the calculated intensities differences, using the Rietveld refinement, considering Si in both $4a$ and $8h$ sublattices.}
	\label{drx-am4}
\end{figure}

\Figs \ref{fig:mev-4}a-b present a comparison of the as-cast and heat-treated Cr$_{0.625}$B$_{0.175}$Si$_{0.2}$ sample. Clearly, during the heat treatment the CrB primary precipitates present in the as-cast sample are consumed, leading to larger amounts of the T1 (Cr$_5$Si$_3$) and T2 (Cr$_5$(Si,B)$_3$) phases in the heat-treated state, as predicted by \citet{Villela2011}. Thus, the heat treatment at 1400°C for 96\,h has led the system very close to thermodynamic equilibrium. Fig. \ref{drx-am4} shows the XRD pattern of the heat-treated sample. All peaks were identified with the T1, T2 and a very small amount of remaining CrB phase not seen in the SEM image. \citet{Nowotny1958} proposed the existence of a D8$_8$ ternary phase at 1300°C, which was contested in the experimental investigations performed by \citet{Villela2011} and \citet{Chad2008} at 1200°C, as the graphite crucible and \edited{low-purity} boron (96.5~at\%) used in \Ref \cite{Nowotny1958} are suspected to stabilize the ternary phase. The present result establishes that the real equilibrium is the Cr$_5$Si$_3$--Cr$_5$(Si,B)$_3$--CrB tie-triangle, showing, also at 1400°C, that the D8$_8$ phase is not stable in the Cr--Si--B system.

Site occupancies of Si and B were refined using the Rietveld method without any constraints on the composition. The only constraint was the full occupancy of the two sites ($4a$ and $8h$) possibly occupied by these two atoms. Therefore, besides the lattice constants, two parameters were refined, with $\chi^2=3.9$ and $R_B=16.1\%$. The Si occupancy on sites $4a$ and $8h$ is found to be \edited{74(2) and 4(2)\%}, respectively, indicating a very strong preference of Si for site $4a$, with only a small fraction going to $8h$. These values correspond to 2.96 Si atoms in $4a$ and 0.32 in $8h$, which are the values shown in Table \ref{tab:defects}. The refined composition was therefore $x_\text{Si}=0.102$, in agreement with \citet{Villela2011}, who suggested a solubility of 9\,at.\%\,Si at 1400°C in the T2 phase. The Rietveld plot is shown in \Fig \ref{drx-am4}.

In order to \edited{support the experimental evidence, we performed \emph{ab-initio} calculations considering} the complete substitution of B by Si in the $4a$ and $8h$ sublattices, resulting in the ordered compounds and a few disordered compounds shown in \Tab \ref{tab:defects}, with mixing of Si and B in the same sublattice. The DFT \cite{Hohenberg1964,Kohn1965} calculations were then performed with the Quantum \textsc{Espresso} code \cite{Giannozzi2009}. The Generalized Gradient Approximation (GGA) was used for the exchange and correlation (XC) functional with the Perdew-Burke-Ernzerhof (PBE) parametrization \cite{Perdew1986}, \edited{considered appropriate for this class of materials} \cite{Medasani2015, Janthon2014, Janthon2013}. A plane-wave cutoff energy of 190\,Ry was used for all calculations, with a 14x14x7 Monkhorst–Pack grid \cite{Monkhorst1976}. \edited{All structures were fully relaxed with respect to lattice parameters and internal degrees of freedom, maintaining the space-group symmetry, by imposing a convergence limit of $10^{-5}$\,Ry in energy and $10^{-3}\,$Ry/au{} in interatomic forces for the ground-state.}

\begin{table*}
	\centering
	\caption{\emph{Ab-initio} and experimental formation enthalpies and lattice parameters for compounds with different T2-structure occupations.}
	\label{tab:defects}
	\begin{tabular}{lccccll}
		\toprule
		Compound & $4a$ occ. & $8h$ occ. & $x_\text{Si}$ & $\Delta_f H$ (kJ/mol) & $a$ (nm) & $c$ (nm) \\
		\midrule
		Cr$_{5}$(B,Si)$_3$ (exp.) & 1.04B/2.96Si & 7.68B/0.32Si & 0.102 & --- & \edited{0.56322(1)} & \edited{1.04197(2)} \\
		Cr:(B)$_{4a}$:(B)$_{8h}$ & 4B & 8B & 0 & $-41.37$ & 0.5414 & 1.0023\\
		Cr:(B$_2$Si$_2$)$_{4a}$:(B)$_{8h}$ & 2B/2Si & 4B & 0.0625 & $-39.13$ & 0.5524 & 1.0124 \\
		Cr:(B)$_{4a}$:(B$_6$Si$_2$)$_{8h}$ & 4B & 6B/2Si & 0.0625 & $-28.15$ & 0.5562 & 1.0091 \\
		Cr:(Si)$_{4a}$:(B)$_{8h}$ & 4Si & 8B & 0.125 & $-38.05$ & 0.5627 & 1.0209\\
		Cr:(B)$_{4a}$:(B$_4$Si$_4$)$_{8h}$ & 4B & 4B/4Si & 0.125 & $-15.76$ & 0.5695 & 1.0226\\
		Cr:(Si)$_{4a}$:(B$_6$Si$_2$)$_{8h}$ & 4Si & 6B/2Si & 0.1875 & $-30.25$ & 0.5743 & 1.0313\\
		Cr:(B)$_{4a}$:(Si)$_{8h}$ & 4B & 8Si & 0.25 & $-0.89$ & 0.5932 & 1.0404\\
		Cr:(Si)$_{4a}$:(Si)$_{8h}$ & 4Si & 8Si & 0.375 & $-12.45$ & 0.6078 & 1.0548\\
		\bottomrule
	\end{tabular}
\end{table*}

\begin{figure*}[t!]
	\centering
	\begin{subfigure}{0.5\textwidth}
		\centering
		\includegraphics[width=1.05\linewidth]{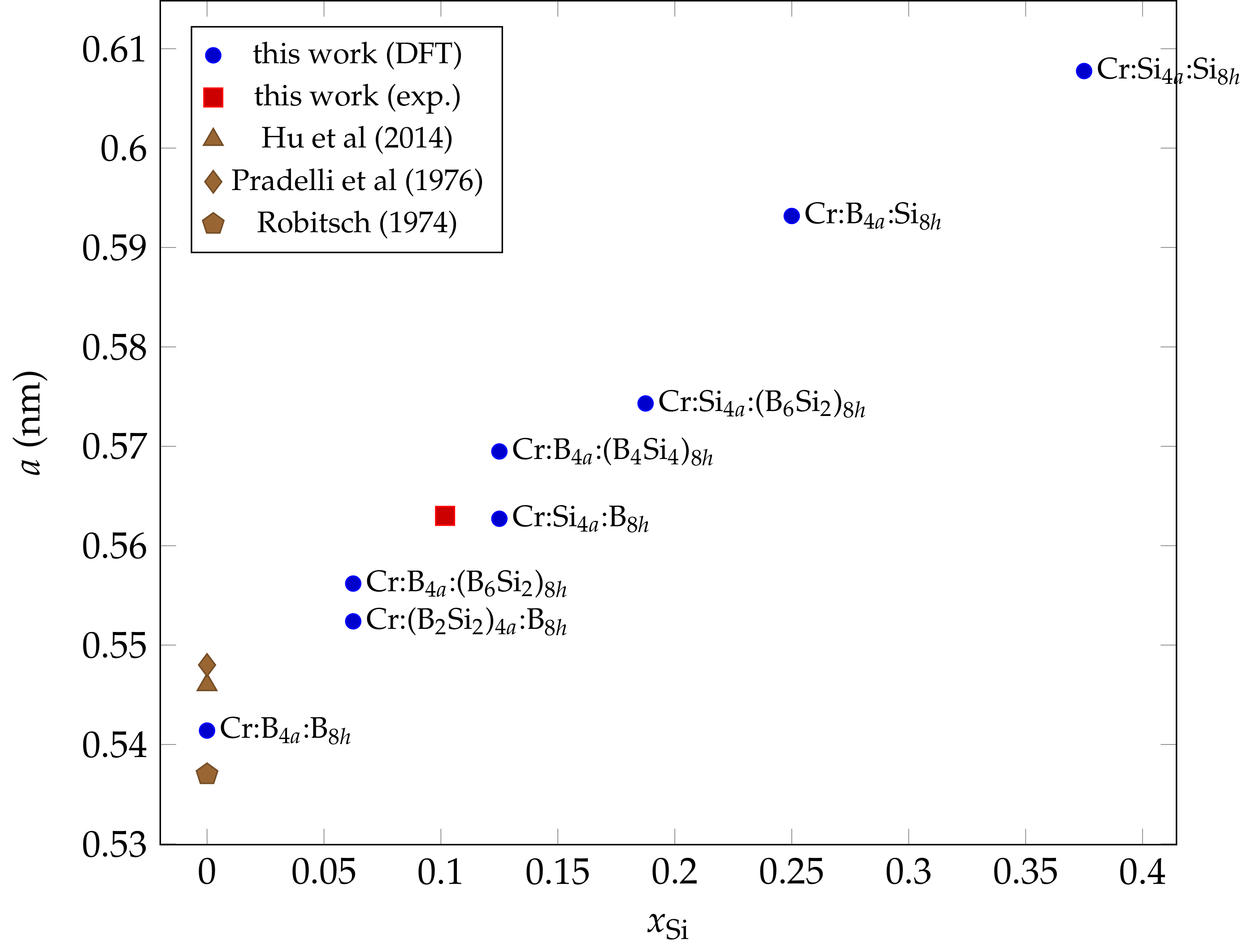}
		\caption{}
	\end{subfigure}%
	\begin{subfigure}{.5\textwidth}
		\centering
		\includegraphics[width=1.05\linewidth]{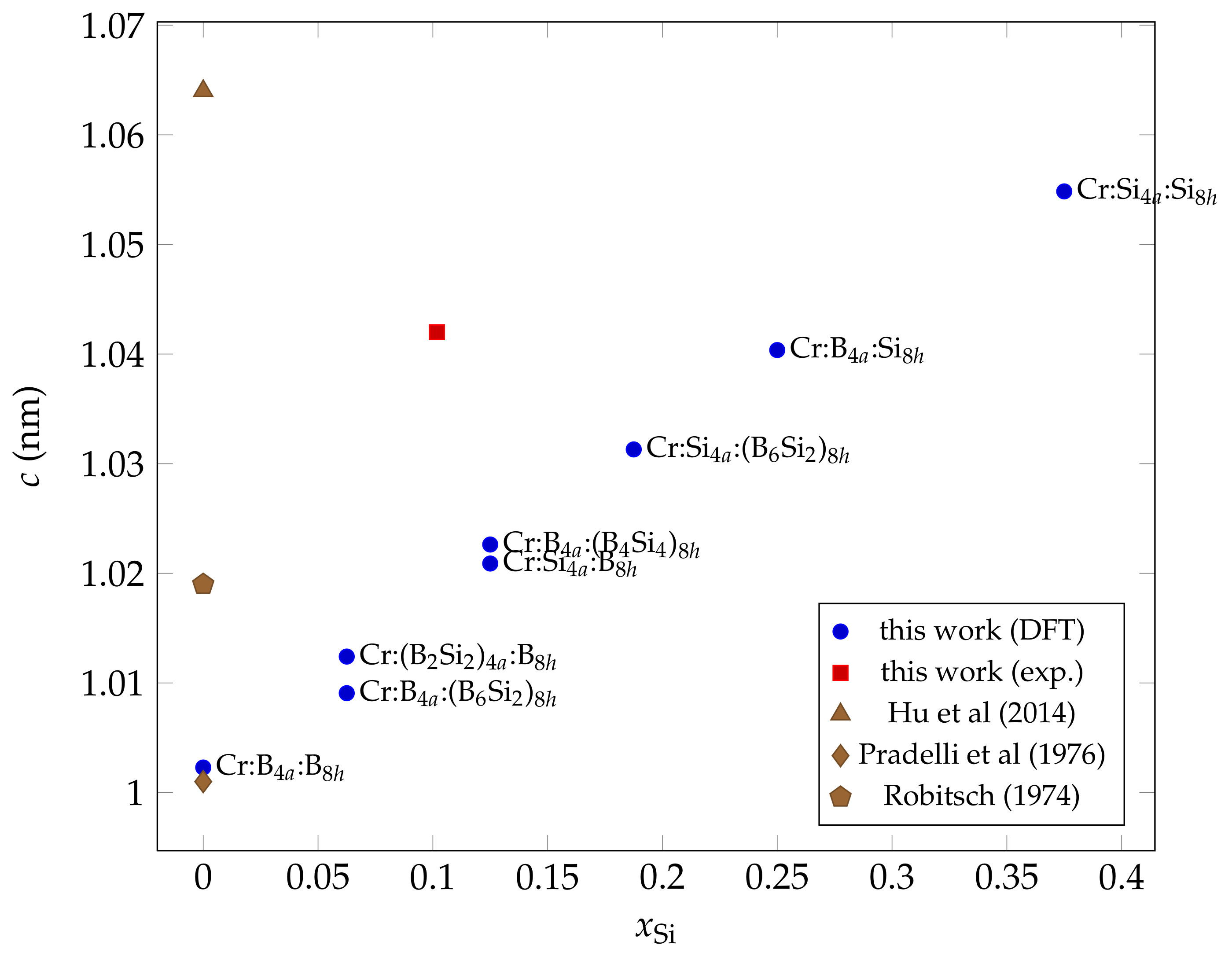}
		\caption{}
	\end{subfigure}
\begin{subfigure}{0.5\textwidth}
	\centering
	\includegraphics[width=\linewidth]{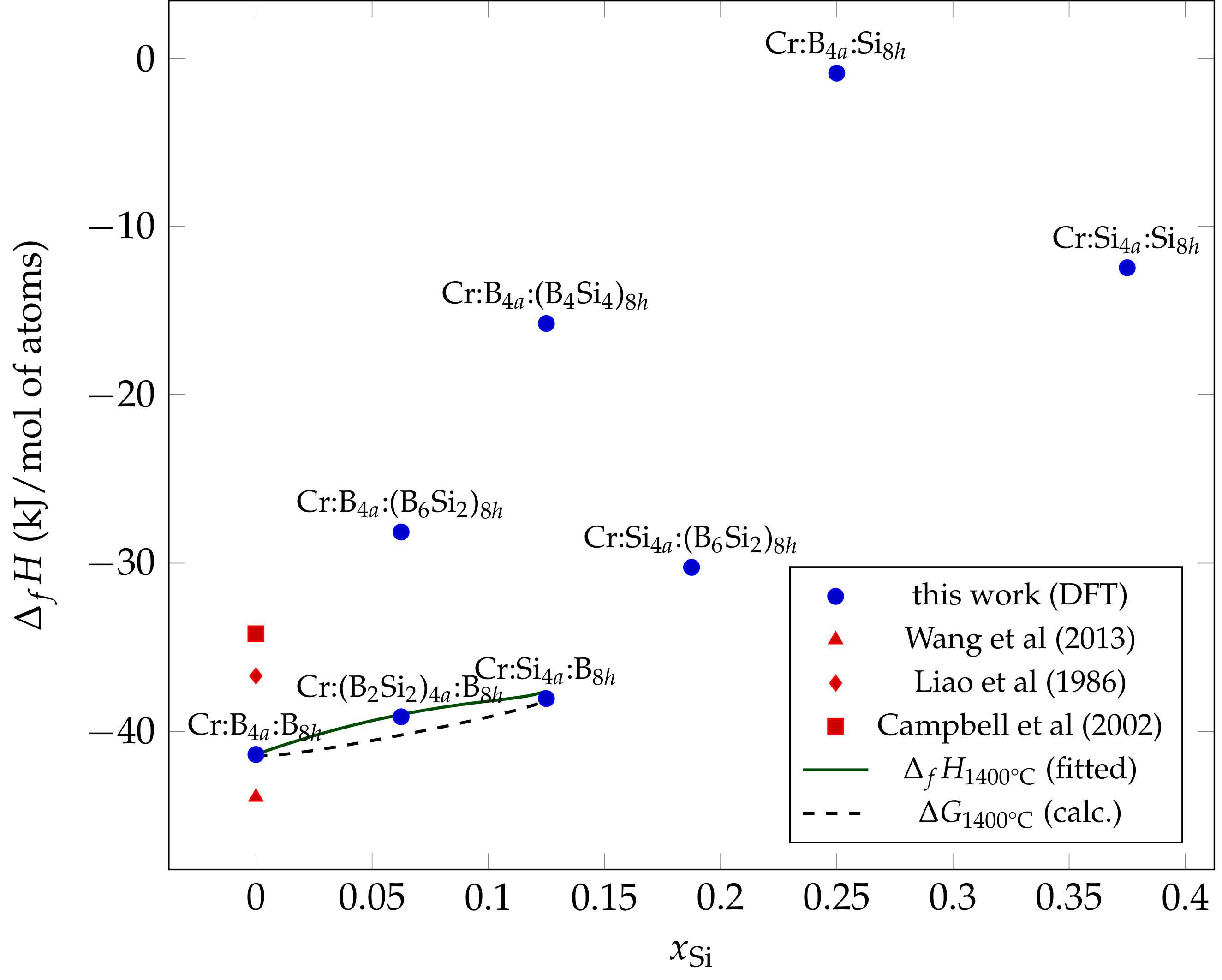}
	\caption{}
\end{subfigure}%
\begin{subfigure}{.5\textwidth}
	\centering
	\includegraphics[width=1.02\linewidth]{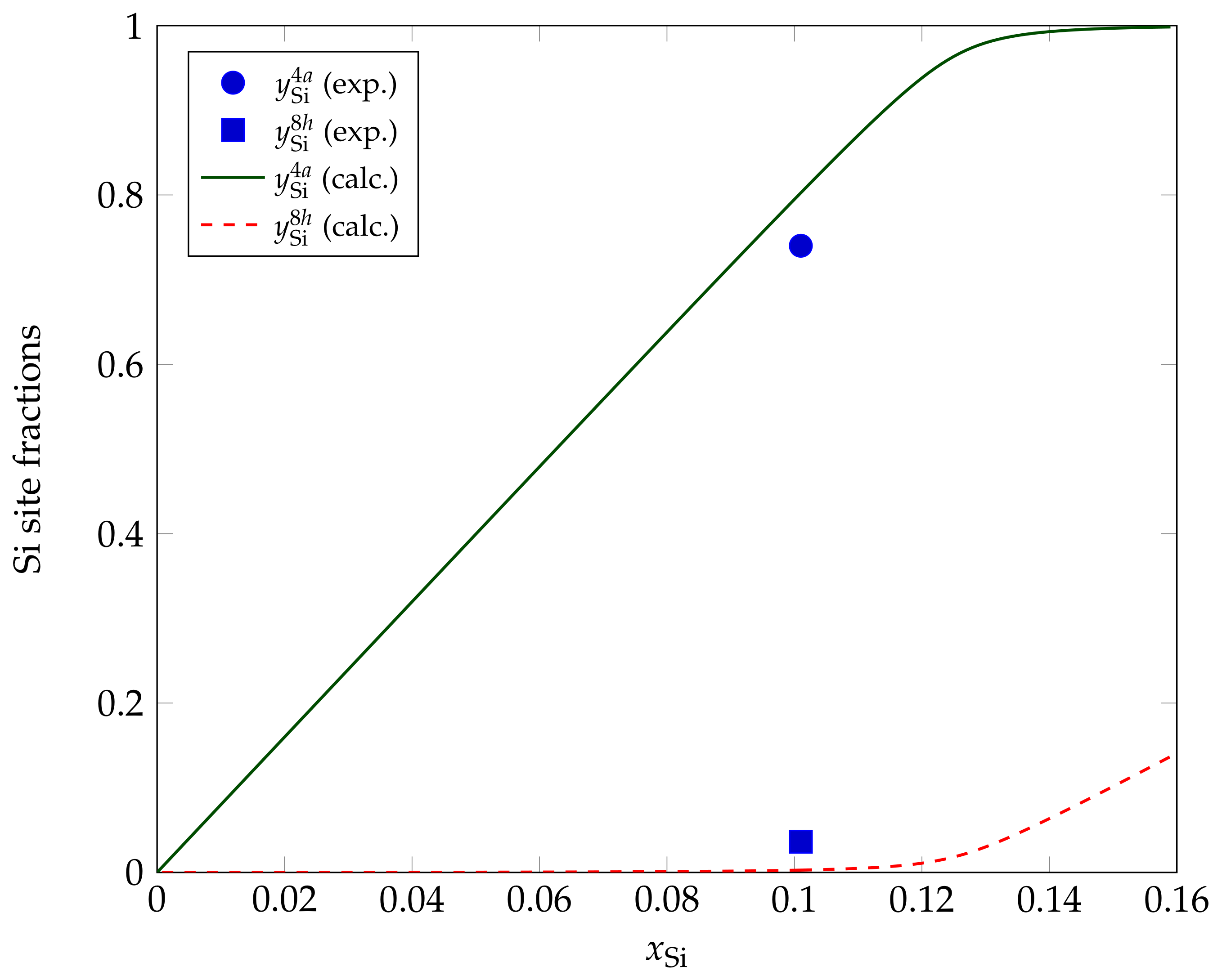}
	\caption{}
\end{subfigure}
	\caption{Comparison between the lattice parameters $a$ (a) and $c$ (b) obtained by DFT calculations, XRD  measurements from synchrotron radiation and results from the literature \cite{Hu2014, Robitsch1974, Pradelli1978}. (c) Comparison between the enthalpies of formation at 0\,K calculated for several T2 configurations and data from the literature (CALPHAD \cite{Campbell2002, Liao1986} and \emph{ab-initio} \cite{Wang2013}) The full line shows the fit using a CALPHAD model at 1400°C and, in dashed lines, the corresponding Gibbs energy curve. (d) Calculated sublattice occupancies using the CALPHAD model and its comparison with the experimental values in \Tab \ref{tab:defects}.}
	\label{fig:info-T2}
\end{figure*}

\edited{Figures \mbox{\ref{fig:info-T2}a-b} present a comparison of experimental and calculated ground-state (0\,K) lattice parameters ($a$ and $c$) obtained in the present work as a function of the Si molar fraction and B and Si site occupancies.} The maximum difference between the results is approximately 2\%, which may be attributed to the XC functional, since the GGA approximation tends to overestimate the forces between atoms, leading to smaller lattice parameters {\cite{Kreutzer2016}}. This difference can be duly noticed in \Figs \ref{fig:info-T2}a-b for both lattice parameters in the case of the Si-free compound (with only one exception for $a$), when compared to the experimental data {\cite{Pradelli1978, Robitsch1974, Hu2014}}.

\Fig \ref{fig:info-T2}c shows the calculated formation enthalpies at 0\,K for all \edited{compounds}, obtained by the difference between the total energy of the \edited{structure} and the total energies of the most stable phases of the pure elements at 0\,K (bcc-A2 Cr, trigonal-$hR12$ B and cubic-A4 Si):
\begin{equation}
	\centering
    \Delta_{\text{f}}{E}_{\text{Cr}_5\text{B}_{3-x}\text{Si}_{x}}= {E}_{\text{Cr}_5\text{B}_{3-x}\text{Si}_{x}}-{5}{E}_\text{Cr}-(3-x){E}_\text{B}-x{E}_\text{Si}
\end{equation}
The value of $-41.4$\,kJ/mol for the binary Cr$_5$B$_3$ compound is very close to those reported by other \emph{ab-initio} \cite{Wang2013} and CALPHAD extrapolation \cite{Liao1986, Campbell2002} results.  At $x_\text{Si}=0.0625$ and $0.125$, the enthalpy values for $4a$ substitution of B by Si are systematically lower than for $8h$. This result is a clear evidence that Si atoms occupy preferentially the $4a$ positions.
%
%

\edited{Some other studies discuss the site occupancies of T2 compounds, as in V--Si--B \mbox{\cite{Nunes2009, Colinet2014}}, Ta--Si--B \mbox{\cite{Khan2012}}, and  Nb--Si--B {\cite{Nunes2012}}. In particular, the latter study \mbox{\cite{Nunes2012}} arrived at the conclusion that B in Nb--Si--B tends to form \emph{dumbbell} bonds in the $8h$ nearest-neighbor positions.} Furthermore, as observed in B$_{14}$Ga$_{3}$Ni$_{27}$ {\cite{Tillard2011}} and Nb$_2$OsB$_2$ {\cite{Mbarki2013}}, for instance, boron atoms have an energetic preference to form short bonds, with interatomic distances varying from 1.40 to 1.90\,\AA, and can be organized, preferentially, in linear or zigzag infinite chains. In the present work, \edited{the \emph{dumbbell} B bonds in the $8h$ sublattices, as shown in Figure {\ref{fig:tab:structure-t2}}}, range from 1.68 to 1.82\,\AA, which are within the expected range. \edited{Furthermore, since the \emph{dumbbell} bonds are very stable, only $4a$ sites remain available for Si occupation. Meanwhile, nearest-neighbor B atoms in $4a$ sites are weakly bonded, at a distance of approximately 3.8\,\AA. That fact allows the partial occupation of these positions by Si.}

In order to estimate the distribution of Si and B among the sublattices from the DFT calculations, we used Thermo-Calc {\cite{Andersson2002}} to calculate the Gibbs energy  at 1400°C, employing the enthalpies of formation of the ordered compounds. \edited{Within the Compound Energy Formalism (CEF) {\cite{Hillert2001}}, we adopted the sublattice model (Cr)$_{5}$:(B, Si)$_{2}$:(B, Si)$_{1}$, the last two sublattices corresponding to $8h$ and $4a$, respectively}. We used an excess term given by $y_\text{B} \, y_\text{Si} \, L$, where $y_i$ is the site fraction of element $i$ in the $4a$ sublattice and $L$ is a fitting parameter found to be equal to 2808\,J/mol of atoms. \edited{The enthalpy fit at 1400°C is shown as a solid line in Fig. {\ref{fig:info-T2}(c)}, whereas the dashed line corresponds to the Gibbs energy. Both curves are shown only for $x_\text{Si}<0.125$, that is, up to full occupation of $4a$ by Si.} \edited{Finally,} Fig. {\ref{fig:info-T2}}(d) shows the $8h$ and $4a$ calculated site occupancies by Si. We obtain therefore a quantitative calculation of Si occupancies showing the preferential nature of Si distribution, showing a reasonable consistency with the Rietveld refined site occupancies, also shown in Fig. {\ref{fig:info-T2}}(d).

In summary, the site occupancies of the T2 phase in the Cr--Si--B system were successfully determined. Synchrotron X-ray diffraction experiments showed that Si occupies preferentially the $4a$ sites. DFT calculations also pointed to the same conclusion. Moreover, as the sublattice model for a phase in a thermodynamic database must be compatible with other systems, we propose the model (Cr)$_5$(B, Si)$_2$(B, Si)$_1$ for the T2 phase in future thermodynamic assessments of relevant systems.

\section*{Acknowledgements}

This study was financed in part by the Coordenação de Aperfeiçoamento de Pessoal de Nível Superior (CAPES) - Brasil - Finance Code 001. The \edited{financial} support of the Fundação de Amparo à Pesquisa do Estado de São Paulo (FAPESP) under Grant No. 2018/10835-6, 2020/08258-0 and 2019/05005-7 is gratefully acknowledged. Finally, the authors acknowledge the Brazilian Synchrotron Light Laboratory (LNLS), Campinas (SP), Brazil, for the high-resolution synchrotron powder diffraction measurements under Project No. 20180142.


\begin{singlespace}
\bibliographystyle{scriptamat}
\bibliography{Cr-Si-B}
\end{singlespace}

\end{document}